\documentclass[11pt]{article}
\usepackage{moriond,epsfig}

\def\MyPhoto{\psfig{figure=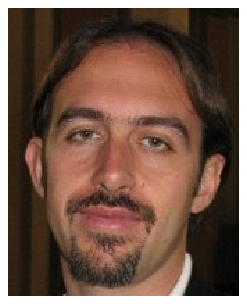,width=35mm}}

\def\be{\begin{equation}}
\def\ee{\end{equation}}
\def\bea{\begin{eqnarray}}
\def\eea{\end{eqnarray}}

\begin{document}
\vspace*{4cm}
\title{The PAMELA Space Experiment}
\author{E. Mocchiutti$^{1}$, O. Adriani$^{2,3}$, G. C. Barbarino$^{4,5}$,
G. A. Bazilevskaya$^{6}$, R. Bellotti$^{7,8}$, M. Boezio$^{1}$,
E. A. Bogomolov$^{9}$, L. Bonechi$^{2,3}$, M. Bongi$^{3}$,
V. Bonvicini$^{1}$, S. Borisov$^{11,14}$, S. Bottai$^{3}$, A. Bruno$^{7,8}$,
F. Cafagna$^{7}$, D. Campana$^{5}$, R. Carbone$^{5,13}$, P. Carlson$^{10}$,
M. Casolino$^{11}$, G. Castellini$^{12}$, M. P. De Pascale$^{11,13}$,
N. De Simone$^{11,13}$, V. Di Felice$^{11,13}$,  
A. M. Galper$^{14}$, W. Gillard$^{10}$, 
L. Grishantseva$^{14}$, P. Hofverberg$^{10}$, G. Jerse$^{1}$,
S. V. Koldashov$^{14}$, S. Y. Krutkov$^{9}$, A. N. Kvashnin$^{6}$,
A. Leonov$^{14}$, O. Maksumov$^{6}$,
V. Malvezzi$^{11}$, L. Marcelli$^{11}$, W. Menn$^{15}$,
V. V. Mikhailov$^{14}$,  N. N. Nikonov$^{11,9}$,
G. Osteria$^{5}$, P. Papini$^{3}$, M. Pearce$^{10}$,
P. Picozza$^{11,13}$, M. Ricci$^{16}$, S. B. Ricciarini$^{3}$, L. Rossetto${^10}$, 
M. Runtso$^{14}$,
M. Simon$^{15}$, R. Sparvoli$^{11,13}$, P. Spillantini$^{2,3}$,
Y. I. Stozhkov$^{6}$, A. Vacchi$^{1}$,
E. Vannuccini$^{3}$, G. Vasilyev$^{9}$, S. A. Voronov$^{14}$, J. Wu${10}$, 
Y. T. Yurkin$^{14}$, G. Zampa$^{1}$, N. Zampa$^{1}$ and
V. G. Zverev$^{14}$ }
\address{1. INFN, Sezione di Trieste, Padriciano 99, I-34012 Trieste, Italy 
 \\ 2. University of Florence, Department of Physics, 
Via Sansone 1, I-50019 Sesto Fiorentino, Florence, Italy
 \\ 3. INFN, Sezione di Florence,  
Via Sansone 1, I-50019 Sesto Fiorentino, Florence, Italy
 \\ 4. University of
Naples 
``Federico II'', Department of Physics, Via Cintia, I-80126 Naples, Italy
 \\ 5. INFN, Sezione di Naples, Via Cintia, I-80126 Naples, Italy
 \\ 6. Lebedev Physical Institute, Leninsky Prospekt 53, RU-119991
Moscow, Russia
 \\ 7 .University of Bari, Department of Physics, Via 
Amendola 173, I-70126 Bari, Italy
 \\ 8. INFN, Sezione di Bari, Via 
Amendola 173, I-70126 Bari, Italy
 \\ 9. Ioffe Physical Technical Institute, Polytekhnicheskaya
26, RU-194021 St. 
Petersburg, Russia
 \\ 10. KTH, Department of Physics, AlbaNova University Centre,
SE-10691 Stockholm, Sweden
 \\ 11. INFN, Sezione di Roma ``Tor Vergata'', Via della Ricerca Scientifica
1, I-00133 Rome, Italy
 \\ 12. IFAC, Via Madonna del Piano 10, I-50019 Sesto Fiorentino,
Florence, Italy
 \\ 13. University of Rome ``Tor Vergata'', Department of Physics, Via
della Ricerca Scientifica 
1, I-00133 Rome, Italy
 \\ 14. Moscow Engineering and Physics Institute, Kashirskoe
Shosse 31, RU-11540 Moscow, Russia
 \\ 15. Universit\"{a}t Siegen, D-57068 Siegen, Germany
 \\ 16. INFN, Laboratori Nazionali di Frascati, Via Enrico Fermi 40,
I-00044 Frascati, Italy}

\maketitle\abstracts{
The 15$^{th}$ of June 2006, the PAMELA satellite--borne experiment was launched from the Baikonur cosmodrome and it has been collecting data since July 2006. The apparatus comprises a time--of--flight system, a silicon--microstrip magnetic spectrometer, a silicon--tungsten electromagnetic calorimeter, an anticoincidence system, a shower tail counter scintillator and a neutron detector. The combination of these devices allows precision studies of the charged cosmic radiation to be conducted over a wide energy range (100 MeV -- 100's GeV) with high statistics. The primary scientific goal is the measurement of the antiproton and positron energy spectrum in order to search for exotic sources, such as dark matter particle annihilations. PAMELA is also searching for primordial antinuclei (anti--helium) and testing cosmic--ray propagation models through precise measurements of the anti--particle energy spectrum and precision studies of light nuclei and their isotopes. Moreover, PAMELA is investigating phenomena connected with solar and earth physics.
}

\section{ Introduction } 
The main stream of physics goals the WiZard Collaboration is devoted to is the study of cosmic rays through balloon and satellite borne devices.

The determination of the antiproton~\cite{boezioA} and positron~\cite{boezioB} spectra, the search of antimatter, the measurement of low energy trapped and solar cosmic rays was carried out using both balloon and satellite experiments (NINA-1~\cite{bidoliA} and NINA-2~\cite{bidoliB}). Other research on board Mir and International Space Station has involved the measurement of the radiation environment, the nuclear abundances and the investigation of the Light Flash phenomenon with the Sileye experiments~\cite{caso1,caso2}.

PAMELA is a dedicated satellite borne experiment conceived by the WiZard collaboration to study the anti--particle component of the cosmic radiation. At present PAMELA is the largest and most complex device built by our collaboration. In this work we describe the scientific objectives, the detector and the first preliminary results of PAMELA after almost three years of data taking.

\section{Physics goals and instrument description}
PAMELA aims to measure in great detail the cosmic ray component at 1 AU (Astronomical Unit). Its 70 degrees, 350--610 km quasi--polar elliptical orbit makes it particularly suited to study items of galactic, heliospheric and trapped nature. PAMELA has been mainly conceived to perform high--precision spectral measurement of antiprotons and positrons and to search for antinuclei, over a wide energy range. Besides the study of cosmic antimatter, the instrument setup and the flight characteristics allow many additional scientific goals to be pursued. Due to the high--identification capabilities of the instrument light nuclei and their isotopes, as well, at least up to Z=8, can be identified. This provides complementary data, besides antimatter abundances, to test models for the origin and propagation of galactic cosmic rays. In addition, the low--cutoff orbit and long--duration mission permits to detect low--energy particles (down to 50 MeV) and to follow long--term time variations of the radiation intensity and transient phenomena. This allows to extend the measurements down to the solar--influenced energy region,  providing unprecedented data about spectra and composition of solar energetic particles and allowing to study solar modulation of galactic cosmic rays over the minimum between solar cycles 23 and 24. Finally, the satellite orbit spans over a significantly large region of the Earth magnetosphere, making possible to study its effect on the incoming radiation. A more detailed overview of the PAMELA scientific goal can be found in~\cite{pamela2}. 

The instrument is installed inside a pressurized container (2~mm aluminum window) attached to the Russian Resurs--DK1 Earth--observation satellite that was launched into Earth orbit by a Soyuz--U rocket on June 15$^{th}$ 2006 from the Baikonur cosmodrome in Kazakhstan. The mission is foreseen to last till at least December 2011.

\begin{figure}[ht] 
\begin{center}
  \includegraphics[width=0.45\textwidth]{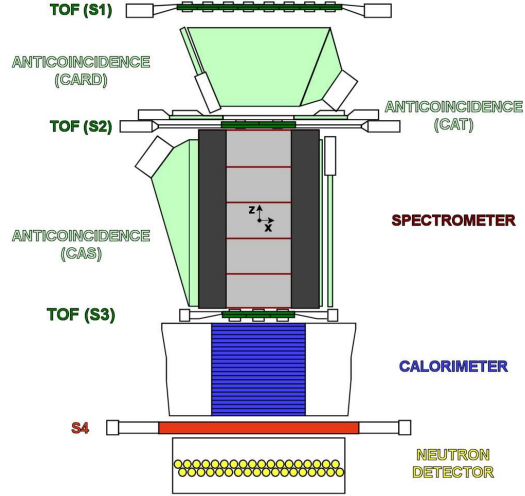}
  \caption{A schematic view of the PAMELA apparatus. The instruments is $\sim$1.3~m tall and has a mass of 470~kg. The average power consumption is 355~W. Magnetic field lines are oriented parallel to the y direction.\label{pamela}}   
\end{center}
\end{figure}
The PAMELA apparatus comprises the following subdetectors, arranged as shown in Figure~\ref{pamela} (from top to bottom): a time--of--flight system (TOF -- S1, S2, S3); a magnetic spectrometer; an anticoincidence system (AC -- CARD, CAT, CAS); an electromagnetic imaging calorimeter; a shower tail catcher scintillator (S4) and a neutron detector. Planes of plastic scintillator mounted above and below the spectrometer form the TOF system which also provides a fast signal for triggering the data acquisition. The timing resolution of the TOF system allows albedo--particle identification and mass discrimination below ~1 GeV/c. The central part of the PAMELA apparatus is the magnetic spectrometer consisting of a 0.43~T permanent magnet and a silicon tracking system, composed of 6 planes of double--sided microstrip sensors. The spectrometer measures the rigidity (momentum over charge) of charged particles and the sign of their electric charge through their deflection (inverse of rigidity) in the magnetic field. Ionization losses are measured in the TOF scintillator planes, the silicon planes of the tracking system and the first silicon plane of the calorimeter allowing the absolute charge of traversing particles to be determined. The acceptance of the spectrometer, which also defines the overall acceptance of the PAMELA experiment, is 21.5~cm$^{2}$sr and the spatial resolution of the tracking system is better than 4~$\mu$m up to a zenith angle of 10$^{\circ}$, corresponding to a maximum detectable rigidity (MDR) exceeding 1~TV. The spectrometer is surrounded by a plastic scintillator veto shield, aiming to identify false triggers and multiparticle events generated by secondary particles produced in the apparatus. Additional information to reject multiparticle events comes from the segmentation of the TOF planes in adjacent paddles and from the tracking system. An electromagnetic calorimeter (16.3~X$_{0}$, 0.6~$\lambda_{0}$) mounted below the spectrometer measures the energy of incident electrons and allows topological discrimination between electromagnetic and hadronic showers, or non--interacting particles. A plastic scintillator system mounted beneath the calorimeter aids the identification of high--energy electrons and is followed by a neutron detection system for the selection of high--energy electrons which shower in the calorimeter but do not necessarily pass through the spectrometer. For this purpose, the calorimeter can also operate in self--trigger mode to perform an independent measurement of the lepton component up to 2~TV.  More technical details about the entire PAMELA instrument and launch preparations can be found in~\cite{pamelone}.

PAMELA was first switched on June 21$^{\mbox{st}}$ 2006 and it has been collecting data continuosly since July 11$^{\mbox{th}}$ 2006. To date about 650 days of data have been analyzed, corresponding to more than one billion recorded triggers and about 12 TB data. 

\section{Anti--particle measurement}

The main task of PAMELA is to identify antimatter components against the most abundant cosmic--ray components. At high energy, main sources of background in the antimatter samples come from spillover (protons in the antiproton sample and electrons in the positron sample) and from like--charged particles (electrons in the antiproton sample and protons in the positron sample). Spillover background comes from the wrong determination of the charge sign due to measured deflection uncertainty; its extent is related to the spectrometer performances and its effect is to set a limit to the maximum rigidity up to which the measurement can be extended. The like--charged particle background is related to the capability of the instrument to perform electron--hadron separation.

\subsection{ Antiproton to proton ratio }

Electrons in the antiproton sample can be easily rejected by applying conditions on the calorimeter shower topology, while the main source of background comes from spillover protons. In order to reduce the spillover background and accurately measure antiprotons up to the highest possible energy, strict selection criteria were imposed on the quality of the fit. To measure the antiproton--to--proton flux ratio the different calorimeter selection efficiencies for antiprotons and protons were estimated. The difference is due to the momentum dependent interaction cross sections for the two particles. These efficiencies were studied using both simulated antiprotons and protons, and proton samples selected from the flight data. In this way it was possible to normalize the simulated proton and therefore the antiproton selection efficiency. The selected proton and antiproton samples could be contaminated by pions produced by cosmic--ray interactions with the PAMELA payload. This contamination was studied using both simulated and flight data. At low energy, below 1~GV, negatively and positively--charged pions were identified in the flight data using the velocity measurement of the ToF system once the calorimeter rejected electrons and positrons from the sample. Since the majority of these pion events are produced locally in the PAMELA structure or pressure vessel it was possible to reject most of them by the mean of strict selection criteria on the AC scintillators and on the energy deposits in either S1 or S2. The energy momentum spectrum of the surviving pions was measured below 1 GV and compared with the corresponding spectrum obtained from simulation. In the simulation, protons impinged isotropically on PAMELA from above and from the side. The protons 
were generated according to the experimental proton spectrum measured by PAMELA and for the pion production both GHEISHA and FLUKA generators~\cite{petter,alessandro} were considered. The comparison between simulated and experimental pion spectrum below 1~GV resulted in a normalization factor for the simulation which accounted for all uncertainties related to the pion production and hadronic interactions. This procedure allowed to estimate the residual pion contamination on all the energy range, resulting to be less than 5\% above 2~GV decreasing at less than 1\% above 5~GV. 

\begin{figure}[!ht]
\begin{center}
  \includegraphics[width=0.55\textwidth]{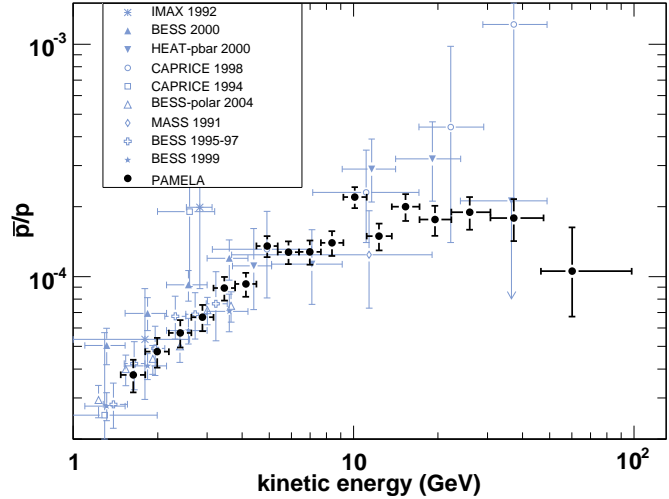} 
  \caption{The antiproton-to-proton flux ratio obtained by PAMELA compared with contemporary measurements.\label{pbar} }
\end{center}
\end{figure}
Figure~\ref{pbar} shows the antiproton--to--proton flux ratio measured by the PAMELA experiment~\cite{aparticle} compared with other contemporary measurements~\cite{boezio1,beach,hof,mitchell,boezio2,asaoka,hams}. Only statistical errors are shown since the systematic uncertainty is less than a few percent of the signal, which is significantly lower than the statistical uncertainty. The PAMELA data are in excellent agreement with recent data from other experiments, the antiproton--to--proton flux ratio increases smoothly with energy up to about 10 GeV and then levels off. The data follow the trend expected from secondary production calculations and our results are sufficiently precise to place tight constraints on secondary production calculations and contributions from exotic sources, e.g. dark matter particle annihilations.

\subsection{Positron fraction}
The main issue in the positron measurement is to control the proton contamination in the positron sample. Two different methods have been used to deal with the proton contamination: the proton background suppression and the proton background estimation.

The first method assume that the proton contamination in the positron sample can be reduced to a negligible amount using strong selection criteria on the topology of the shower inside the calorimeter and by requiring the match between the calorimeter detected energy and the tracker measured momentum. Using particle beam data collected at CERN we have previously shown~\cite{boe06} that a proton rejection power of 10$^5$ can be achieved up to 200 GeV/c keeping an electron selection efficiency of 80\%. 

The second approach consists in keeping a very high selection efficiency and in quantifying the residual proton contamination by the mean of a so--called ``spectral analysis''~\cite{psarticle}. In a conservative approach, the proton distributions needed to estimate the contamination were obtained using the flight calorimeter data without any dependence on simulations or test beam data. The calorimeter was divided in two parts: the upper part (``pre--sampler'')  made of two tungsten planes and four detector planes was used to reject non-interacting particles. The sample of events passing this condition is a nearly pure sample of protons with a positron contamination of less than 2\% at rigidities greater than 1.5~GV. Calorimeter variables were evaluated for the lower part of the calorimeter made of 20 tungsten planes and 40 detector planes and the distribution of the lateral shower spread for protons was obtained. Hence, positive particles were selected using the first 20 tungsten planes of the calorimeter.
\begin{figure}[!ht]
\begin{center}
  \includegraphics[width=0.7\textwidth]{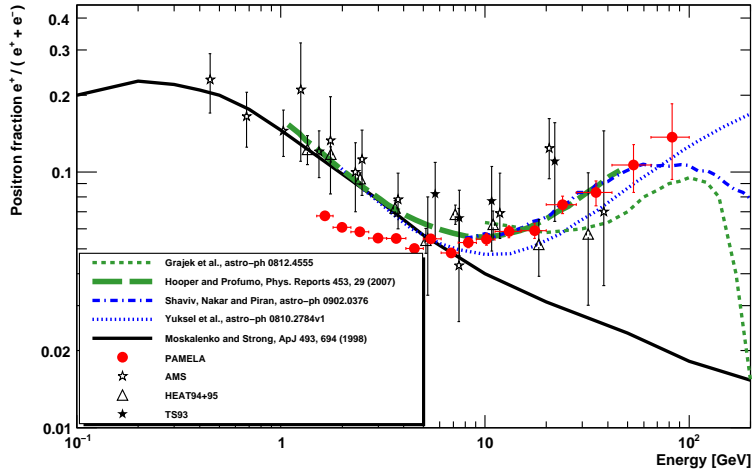}
  \caption{The PAMELA positron fraction compared to other experimental results and theoretical predictions.\label{psratio}}
\end{center}
\end{figure}
Results are shown in figure~\ref{psratio}, where PAMELA data~\cite{psarticle} are compared to some contemporary measurements~\cite{goldenps,alcaraz,beatty} and to theoretical predictions for dark matter~\cite{hooper,grajek} and astrophysical~\cite{shaviv,yuksel} primary production. At low energy PAMELA data are lower than most of the other data and this can be interpreted as an observation of charge-sign dependent solar modulation effects. PAMELA data are in agreement with results from a balloon--borne experiment which flew in June 2006~\cite{clem} which observed a low positron fraction at low energies, but with large statistical uncertainties. At higher energies and below 10 GeV the PAMELA positron fraction is compatible with other measurements but does not confirm the structure between 6 and 10 GeV which was claimed previously by the HEAT experiment~\cite{cou99}. Above 10 GeV the positron fraction is in agreement with the recent measurements and it increases significantly with energy. The PAMELA data cannot be described by the standard model of secondary production, black line~\cite{moskalenko} in figure~\ref{psratio}.

\section{Other physics results}

\subsection{Galactic cosmic rays}
\begin{figure}[!t]
  \centering
  \begin{minipage}[t]{0.49\textwidth}
    \centering
    \includegraphics[width=0.90\textwidth]{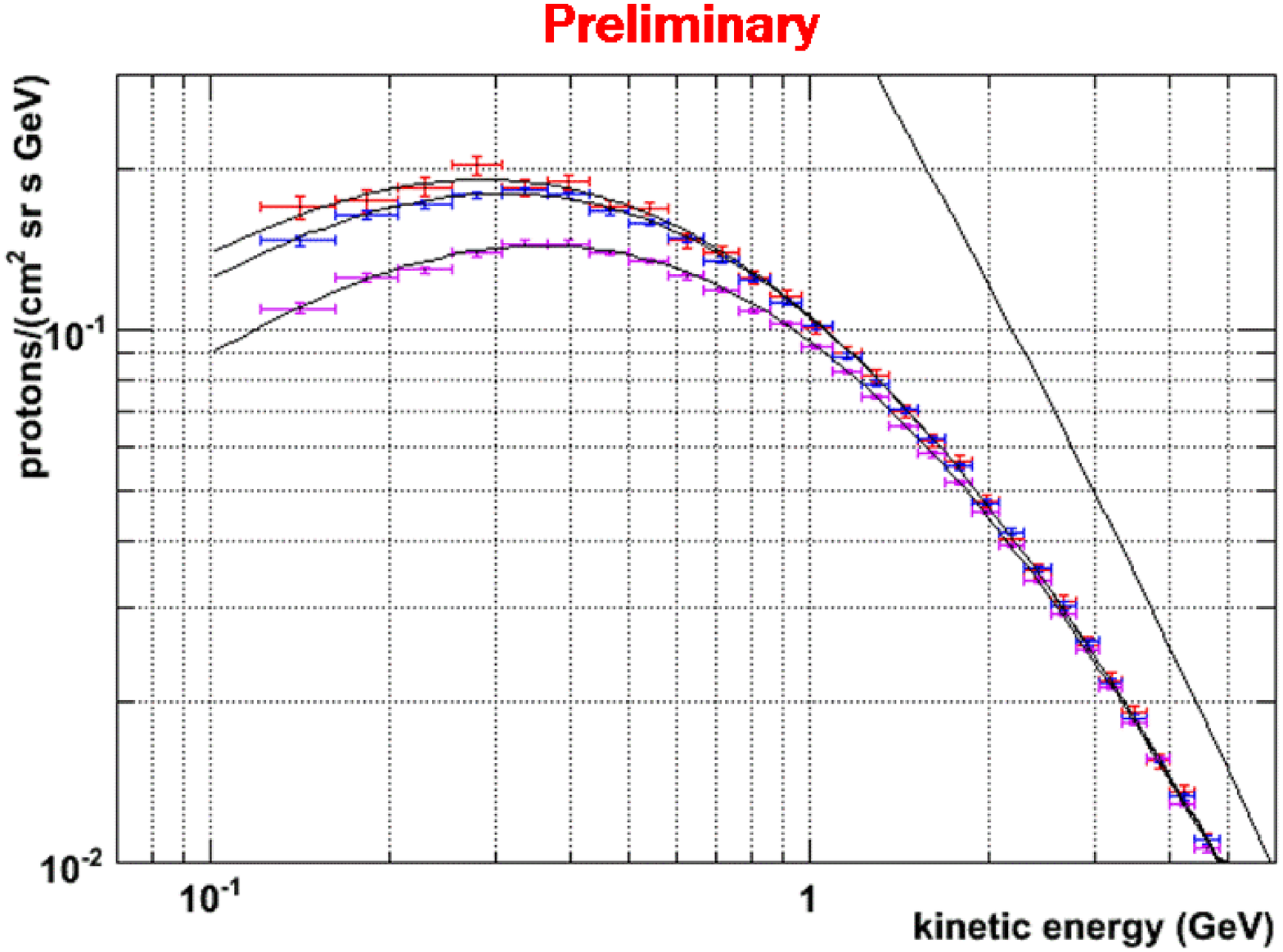}
    \caption{Proton flux measured by PAMELA in July 2006 (violet), August 2007 (blue) and February 2008 (red). Solid line represents the interstellar spectrum.}
    \label{phesolarmod1}
  \end{minipage}
  \hfill
  \begin{minipage}[t]{0.49\textwidth}
    \centering
    \includegraphics[width=\textwidth]{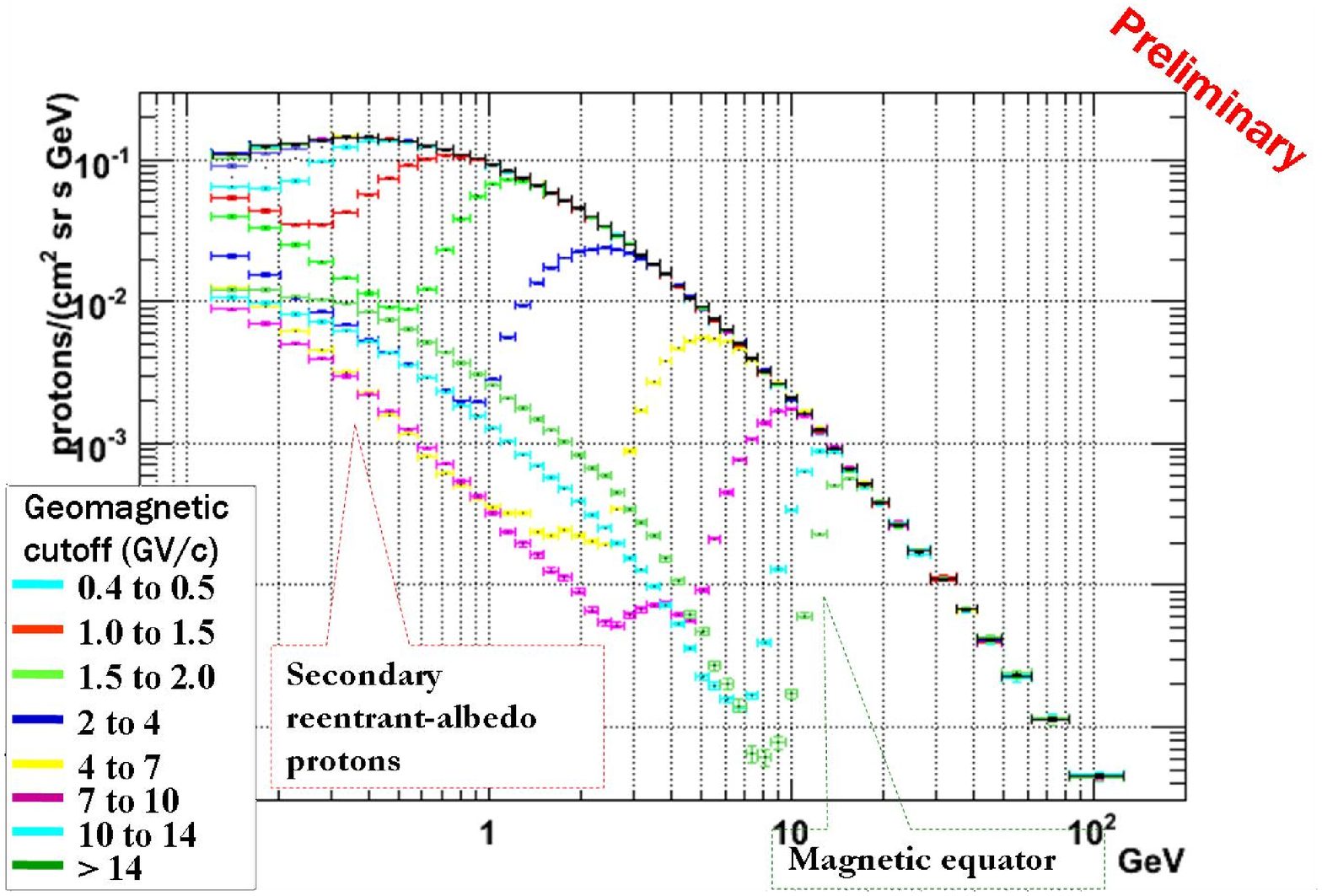}
    \caption{Primary and re-entrant albedo proton spectra as function of the geomagnetic cutoff.}
    \label{phesolarmod2}
  \end{minipage}
\end{figure}

The most common particles in the cosmic radiation, protons and helium nuclei, are detected by PAMELA with very high statistics over a wide energy range. This allows to perform a precise measurements of their spectral shape and makes possible to study time variations and transient phenomena. The measurement of the spectra of primary cosmic rays has deep astrophysical implications and the importance of knowing the absolute values of the fluxes is essential to perform, for example, atmospheric neutrino observations.
The launch of PAMELA occurred during the 23$^{rd}$ solar minimum. In this period it is possible to observe solar modulation of galactic cosmic rays due to varying solar activity. A long term measurement of the proton, electron and nuclear flux at 1 AU provides information on propagation phenomena occurring in the heliosphere. As already mentioned, the possibility to identify the anti--particle spectra allows to study also charge dependent solar modulation effects.
Figure~\ref{phesolarmod1} shows the proton flux as measured by PAMELA in three different months in 2006, 2007 and 2008. An increasing flux of galactic cosmic rays corresponds to a decreasing solar activity. This effect is in
agreement with the increase of neutron monitor fluxes~\cite{neumon}.

\subsection{Nuclei }
Also light nuclei (up to Oxygen) are detectable with the scintillator system. In this way it is possible to study with high statistics the secondary/primary cosmic ray nuclear and isotopic abundances such as B/C, Be/C, Li/C and $^3$He/$^4$He. These measurements can constrain existing production and propagation models in the galaxy, providing detailed information on the galactic structure and the various mechanisms involved. The preliminary B/C ratio as function of kinetic energy per nucleon measured by PAMELA is in very good agreement with previous measurements.

\subsection{Primary, re-entrant albedo and trapped particles measurements}
Albedo particles are secondary particles produced by cosmic--rays interacting with the Earth's atmosphere that are scattered upward. When these particles lack sufficient energy to leave the Earth's magnetic field they re-enter the atmosphere in the opposite hemisphere but at a similar magnetic latitude and they are called re-entrant albedo particles. The measurement of the composition and spectra of the secondary cosmic rays particles provides a tool for the fine tuning of models used in air shower simulation programs. Due to its orbit PAMELA is able to provide a world map of the primary and re-entrant albedo particles, allowing to discern fine details in the spectra especially in the sub-cutoff region. Figure~\ref{phesolarmod2} shows the proton spectra at different geomagnetic latitudes. For each spectra is clearly visible the position of the geomagnetic cutoff and the secondary proton component due to re-entrant albedo particles. It is also possible to observe structures in the spectra, for example the violet data (cutoff 7-10 GV/c) between 3 and 5 GeV. These structures are also reproduced by simulations~\cite{honda}.
The 70$^\circ$ orbit of the Resurs--DK1 satellite allows for continuous monitoring of the electron and proton belts. The high energy ($>$80 MeV) component of Van Allen Belts can 
be monitored during the time and it is possible to perform a detailed mapping of these regions by determining spectral and geometrical features~\cite{casolinowaseda}.

\subsection{ Solar energetic particles }
Due to the period of solar minimum few significant solar events with energy high enough to be detectable are expected. The observation of solar energetic particle (SEP) events with a magnetic spectrometer will allow several aspects of solar and heliospheric cosmic ray physics to be addressed for the first time. Positrons are produced mainly in the decay of $\pi^{+}$ coming from nuclear reactions occurring at the flare site. Up to now, they have only been measured indirectly by remote sensing of the gamma ray annihilation line at 511 keV. Using the magnetic spectrometer of PAMELA it is possible to separately analyze the high energy tail of the electron and positron spectra at 1 AU obtaining information both on particle production and charge dependent propagation in the heliosphere in perturbed conditions of solar particle events. PAMELA is able to measure the spectrum of cosmic ray protons from 80 MeV up to almost 1 TeV and therefore is able to measure the solar component over a very wide energy range (where the upper limit will be limited by statistics). Furthermore PAMELA can also investigate the light nuclear component related to SEP events over a wide energy range. These measurements will help us to better understand the selective acceleration processes in the higher energy impulsive events~\cite{rea}. 

\section{Conclusions}
PAMELA is continuously taking data and the mission is planned to continue until at least December 2011. The increase in statistics will allow higher energies to be studied. An analysis for low energy antiprotons and positrons (down to  100 MeV) is in progress and will be the topic of future publications.

\section*{Acknowledgements}
We would like to acknowledge contributions and support from: Italian Space Agency (ASI), Deutsches Zentrum f\"ur Luft-- und Raumfahrt (DLR), The Swedish National Space Board, Swedish Research Council, Russian Space Agency (Roskosmos, RKA) and russian grants: RFBR grant 07-02-00992a and Rosobr grant 2.2.2.2.8248.

\end{document}